# Insight into the correlation between lag time and aggregation rate in the kinetics of protein aggregation

Stefan Auer[1,*] and Dimo Kashchiev[2]

[1]*Centre for Molecular Nanoscience, University of Leeds, Leeds, LS2 9JT, UK*
[2]*Institite of Physical Chemistry, Bulgarian Academy of Sciences, ul. Acad. G. Bonchev 11, Sofia 1113, Bulgaria*

*Corresponding author.* E-mail address: S.Auer@leeads.ac.uk

**ABSTRACT:** Under favourable conditions, many proteins can assemble into macroscopically large aggregates such as the amyloid fibrils that are associated with Alzheimer's, Parkinson's and other neurological and systemic diseases. The overall process of protein aggregation is characterized by initial lag time during which no detectable aggregation occurs in the solution and by maximal aggregation rate at which the dissolved protein converts into aggregates. In this study, the correlation between the lag time and the maximal rate of protein aggregation is analyzed. It is found that the product of these two quantities depends on a single numerical parameter, the kinetic index of the curve quantifying the time evolution of the fraction of protein aggregated. As this index depends relatively little on the conditions and/or system studied, our finding provides insight into why for many experiments the values of the product of the lag time and the maximal aggregation rate are often equal or quite close to each other. It is shown how the kinetic index is related to a basic kinetic parameter of a recently proposed theory of protein aggregation.

A wide range of different proteins unrelated in their amino acid sequence have been shown to convert into large ordered aggregates known as amyloid fibrils that are associated with various neurological and systemic diseases.[1-3] Amyloid fibrils have common characteristic optical properties such as birefringence on binding of certain dyes including Congo red and thioflavin-T.[1,2] Also, X-ray diffraction experiments revealed that amyloid fibrils share a common cross-β structure formed of intertwined layers of β-sheets that are oriented parallel to the fibril elongation axis.[4,5] Protein aggregates may form in a nucleated-mediated manner,[6-22] and the overall aggregation process is characterized by an initial lag time $t_l$ (s) during which no detectable aggregation occurs and by a maximal aggregation rate $k_a$ (s$^{-1}$) at which the proteins convert into fibrils. Measurements of $t_l$ and $k_a$ have been performed[19-21,23-29] with the aid of various techniques, for instance by monitoring the fluorescence signal arising from the binding of dye molecules to the protein aggregates. In $\alpha$-vs.-$t$ coordinates, the normalized fluorescence signal $\alpha$ that describes the course of the process of protein aggregation with time $t$ often has a sigmoidal shape (see, e.g., Refs. 19, 20). The curve in Fig. 1 shows this signal in overall aggregation of β$_2$-microglobulin (β$_2$m).[19]

There are several ways of defining the lag time and the aggregation rate. For example, in the experiments of Xue et al.[19] and Routledge et al.[20] $k_a$ is taken to be the slope of the linear portion of the $\alpha(t)$ curve, visualized by the dashed line in Fig. 1, and the intercept of this line with the time axis is identified as $t_l$. In general, quantitative measurements of $t_l$ and $k_a$ are challenging, because the stochastic nature of the nucleation process involved in overall aggregation causes the values of these quantities to scatter significantly. Also, it is difficult to remove all pre-aggregated material, and there may be secondary processes such as fragmentation and flocculation that are difficult to control. Nevertheless, in order to investigate the effect of mutations and of the experimental conditions on the protein aggregation kinetics, in recent years numerous experiments were reported with measurements of $t_l$ and $k_a$ for different protein systems, including insulin,[23,28] Alzheimer's Aβ(1-40)[26,27] and glucagon.[29] Although these proteins have no sequence similarity and the conditions under which the experiments were performed differed considerably, a systematic comparison[30] of kinetic data revealed a correlation between the lag time and the aggregation rate. In particular, it was found by Fändrich[30] that the product $t_l k_a$ can be represented as $t_l k_a = a$, and the estimate $a = 4.5$ for the numerical parameter $a$ was obtained from a best fit to a large number of experimental data. Most intriguing in this finding is perhaps that although the individual values of $t_l$ and $k_a$ change strongly with the protein sequence and environmental parameters, the product $t_l k_a$ is much less sensitive to these factors (the standard deviation of the above $a$ value is ±2.9). This fact led Fändrich to the suggestion for "mechanistic similarities in the nucleation behaviour of different amyloid-like fibrils and aggregates."[30]

In this study we use concepts from the theory of overall crystallization (e.g., Ref. 31) in order to describe the kinetics of overall aggregation of proteins into amyloid fibrils or other aggregates. In particular, our aim is to provide insight into the reason for which the product $t_l k_a$ seems to be much less system-specific than the individual values of $t_l$ and $k_a$. For analysis of kinetic $\alpha(t)$ data for overall protein aggregation (see Fig. 1) we propose the use of the following general fitting function:

$$\alpha(t) = 1 - \exp\left[-\left(\frac{t}{\theta}\right)^n\right]. \qquad (1)$$

Here $\alpha$, a number between 0 and 1, denotes any normalized experimental observable that increases with the fraction of protein converted into amyloid fibrils or other aggregates. The normalized fluorescence signal is an example for such an observable. Also, $\theta$ (s) is the aggregation time constant determining the time scale of the overall aggregation process, and $n \geq 1$ is the aggregation kinetic index pointing how long it takes (in units of $\theta$) for the process to be accomplished after the lag time. Geometrically, $n$ characterizes the steepness of the linear portion of the $\alpha(t)$ curve in $\alpha$-vs.-$t/\theta$ coordinates: the greater $n$, the steeper this portion.

The above fitting function is in fact a rather general form of the $\alpha(t)$ function in the Kolmogorov-Johnson-Mehl-Avrami (KJMA) theory of overall crystallization (see, e.g., Ref. 31), in which $\alpha$ is the fraction of the total volume or mass crystallized. We note also that for $t/\theta \ll 1$ Eq. (1) turns into the equation used in Ref. 7 for analyzing the early stage of overall protein aggregation. Model kinetic considerations are necessary to reveal the concrete dependencies of $\theta$ and $n$ on the particular experimental conditions and the physical quantities (such as the rates of nucleation, growth, fragmentation, etc.) characterizing the various processes involved in the overall phase-transformation process. For instance, in order to describe the kinetics of overall crystallization by the polynuclear mechanism when crystallites are continuously nucleated, it is assumed that all crystallites are isomorphic and grow irreversibly without fragmentation until the phase transformation is accomplished. In the special case when the crystallite nucleation and growth rates are time-independent, the kinetic index $n$ is readily obtainable (e.g., Ref. 31) and the result is $n = 4$, 3 or 2 for sphere-like, plate-like or needle-like crystallites, respectively. It should be emphasized, however, that when under the same conditions $\alpha$ is the normalized decrement of the intensity of transmitted light rather than the fraction of volume or mass crystallized, for sphere-like crystallites the kinetic index is found to be $n = 3$ or 7, depending on the regime in which light is scattered by the crystallites.[31] Thus, $n$ can have values in a rather limited range when the crystallites nucleate and grow under different experimental conditions and the crystallization process is monitored by different experimental techniques. In stark contrast to $n$, the time constant $\theta$ varies widely (e.g., from seconds to days) with changing the process conditions, because it is very sensitive to the crystallite nucleation and growth rates.

Using the above fitting function, we define $k_a$ as the maximal rate of aggregation and, hence, determine it from the slope of the $\alpha(t)$ curve at the inflection point $t_0$:

$$k_a = \left(\frac{d\alpha}{dt}\right)_{t=t_0}. \tag{2}$$

Also, the time intercept of the tangent to the $\alpha(t)$ curve at $t_0$ defines the lag time $t_l$, i.e.

$$t_l = t_0 - \frac{\alpha_0}{k_a}, \tag{3}$$

where $\alpha_0 \equiv \alpha(t_0)$ (see Fig. 1). Employing $\alpha$ from Eq. (1) in the condition $(d^2\alpha/dt^2)_{t=t_0} = 0$ for inflection and in Eqs. (2) and (3), we find that ($n \geq 1$)

$$t_0 = \left(\frac{n-1}{n}\right)^{1/n}\theta \tag{4}$$

$$\alpha_0 = 1 - e^{-(n-1)/n} \tag{5}$$

$$k_a = (n-1)\left(\frac{n}{n-1}\right)^{1/n}e^{-(n-1)/n}\frac{1}{\theta} \tag{6}$$

$$t_l = \left(\frac{n-1}{n}\right)^{1/n}\left[1 - \frac{e^{(n-1)/n}-1}{n-1}\right]\theta. \tag{7}$$

As seen, individually, the values of $k_a$ and $t_l$ are functions of both $n$ and $\theta$, whereas the product

$$t_l k_a = a(n), \tag{8}$$

with $a$ given by ($n \geq 1$)

$$a(n) = ne^{-(n-1)/n} - 1, \tag{9}$$

depends solely on the kinetic index $n$. This explains why $t_l$ and $k_a$ vary widely in aggregation of various proteins under different experimental conditions, but their product $a$ remains nearly the same.[30] Indeed, as the aggregation time constant $\theta$ changes strongly with these conditions, so do $k_a$ and $t_l$ from Eqs. (6) and (7). In contrast, since the kinetic index $n$ is not too sensitive to these conditions, $a$ from Eq. (9) alters relatively little. This is seen in Fig. 2 in which the lines depict the $a(n)$ dependence from Eq. (9).

We shall now illustrate this dependence by analyzing a certain number of the many experimental $\alpha(t)$ curves of Xue et al.[19] and Routledge et al.[20] on the fibrillation of $\beta_2$m. In Ref. 19 the concentration effect, and in Ref. 20 the effect of mutations on the aggregation kinetics of this protein were investigated. By fitting Eq. (1) to the experimental $\alpha(t)$ data in the way illustrated in Fig. 1, we first determined $n$ and $\theta$. Then, with the help of Eq. (9) we calculated $a$ at the corresponding $n$. The resulting $a$ values are represented by symbols in Fig. 2 and, as observed, they differ from each other and from Fändrich's value of 4.5.[30] The $a$ values are in the range from 2.0 at $n = 7.2$ to 6.2 at $n = 18.4$. For comparison, Fändrich's $a = 4.5$ corresponds to $n = 14$. As seen in Fig. 2, despite that the $n$ values for different systems and conditions are in different groups, there is overlapping between some of them. This is due to the stochastic nature of the individual $\alpha(t)$ curves belonging to one and the same group.

Analogously to the kinetic index in overall crystallization (e.g., Ref. 31), in overall protein aggregation we may expect $n$ to change relatively little with the experimental conditions and the experimental method (fluorescence, turbidity, etc.) for monitoring the process. In other words, $n$ is quite system-independent, and this explains why very different experimental systems may have the same $a$ value. Indeed, according to Eq. (9), for this to be the case it is only necessary the $\alpha(t)$ curves of these systems to posses the same kinetic index $n$. Figure 3 represents all $t_l, k_a$ data analyzed by Fändrich.[30] The solid line in the figure corresponds to his $a = 4.5$, and the two dashed lines show that the $a$ values describing all $t_l, k_a$ data analyzed by him vary between 0.25 and 24. From Eq. (9) it then follows that this variation is due to values of the kinetic index $n$ between 2.3 and 70. It is interesting that while the former value corresponds to a rather slow rise of $\alpha$ from zero to unity, the latter one corresponds to an almost vertically ascending $\alpha(t)$ curve.

The present study does not provide explanation of the physical nature of the aggregation time constant $\theta$ and kinetic index $n$, for as already noted, revealing this nature requires model kinetic considerations. Such considerations are the basis not only of the KJMA theory of overall crystallization, but also of different theoretical descriptions of the kinetics of overall protein aggregation (e.g., Refs. 6-8,10,13,21,32,33), most of which employ particular forms of the general master equation of first-order phase transitions.[31,34]

Let us now use the expressions

$$k_a = \frac{\kappa}{e} \tag{10}$$

$$t_l = \left[ \ln\left(\frac{1}{C_+}\right) - e + 1 \right] \frac{1}{\kappa} \tag{11}$$

of Knowles et al.[21] in order to see how the aggregation time constant $\theta$ and kinetic index $n$ are related to the kinetic parameters $\kappa$ and $C_+$ of these authors' model of the overall process of protein fibrillation. In Eqs. (10) and (11) $e \approx 2.718$ is the base of the natural logarithm, $\kappa$ (s$^{-1}$) is the rate of multiplication of the fibril population (it is a function of the fibril growth and fragmentation rates), and $C_+ << 1$ is a complex quantity depending on the total protein mass, on the initial number and mass of the fibrils and on the fibril nucleation, growth and fragmentation rates. Multiplying $t_l$ from Eq. (11) by $k_a$ from Eq. (10), comparing the right-hand side of the resulting equation with that of Eq. (8) and using $a(n)$ from Eq. (9), we obtain ( $n \geq 1$ )

$$ne^{1/n} = \ln\left(\frac{1}{C_+}\right) + 1. \tag{12}$$

This transcendental algebraic equation for $n$ represents implicitly the $n(C_+)$ dependence corresponding to Knowles et al.'s model.[21] Explicitly, this dependence is readily obtainable, but only approximately. Indeed, employing the truncated series expansion $e^{1/n} \approx 1 + 1/n$ which is valid for $n > 1$, to a good approximation (the error is less than 13% when $n \geq 2$ ), from Eq. (12) we find that

$$n = \ln\left(\frac{1}{C_+}\right) \tag{13}$$

provided $C_+ < e^{-1} \approx 0.37$ (then $n > 1$ ). For unseeded fibrillation, the expression for $C_+$ in Ref. 21 and the values given there for the quantities determining $C_+$ (see the caption of Fig. 3 in Ref. 21) yield $C_+ = 5 \times 10^{-10}$ to $5 \times 10^{-4}$. With these values of $C_+$ it follows from Eq. (13) that for the $k_a$-vs.-$t_l$ data and the $\alpha(t)$ curves in Fig. 3 of Ref. 21 the kinetic index $n$ ranges from 7.6 to 21. According to Eq. (9), these $n$ values correspond to $a$ between 2.2 and 7.1, i.e. to $a$ comparable with Fändrich's $a = 4.5$.[30] As to the dependence of $\theta$ on $\kappa$ and $C_+$, setting equal the right-hand sides of Eqs. (6) and (10) results in the expression ( $n \geq 1$ )

$$\theta = \left(\frac{en}{n-1}\right)^{1/n} \frac{n-1}{\kappa} \tag{14}$$

in which $C_+$ is implicit via $n$ from Eq. (12). With increasing $n$ the factor $[en/(n-1)]^{1/n}$ in this expression tends to unity and can be omitted. Thus, using $n$ from Eq. (13), we find that for $C_+ < e^{-1} \approx 0.37$ (then $n > 1$ ), the explicit dependence of $\theta$ on $\kappa$ and $C_+$ is approximately given by

$$\theta = \left[\ln\left(\frac{1}{C_+}\right) - 1\right]\frac{1}{\kappa}, \tag{15}$$

the error being less than 27% when $C_+ \leq e^{-4} \approx 0.018$ (then $n \geq 4$). Comparing Eqs. (13) and (15), we see that while the particular protein system and aggregation conditions affect relatively little the kinetic index $n$ (only logarithmically via $C_+$), they exert a much stronger influence on the time constant $\theta$ because of the inverse proportionality of $\theta$ to the rate $\kappa$ of multiplication of the fibril population.

In conclusion, the above analysis shows that, strictly, the value $a$ of the $t_i k_a$ product is not a universal number independent of the various protein systems and experimental conditions. Nonetheless, as found by Fändrich,[30] $a$ values around 4.5 are most likely to characterize the process of overall aggregation of proteins. Through Eq. (9), these values contain information about the kinetic index $n$ of the process and, as indicated by their standard deviation, they are predominantly in the range from 1.6 to 7.4. According to Eq. (9), values of $a$ in this range correspond to $n$ values between 6 and 22, and Fändrich's $a = 4.5$ corresponds to $n = 14$. Model considerations are required to reveal the physical nature of both the kinetic index $n$ and the time constant $\theta$ of the overall process of protein aggregation. Equations (12) and (14) reveal this nature in the scope of Knowles et al.'s theory[21] of the kinetics of overall fibrillation. For example, if we apply this theory to Xue et al.'s experiments,[19] with the respective $n$ and $\theta$ values of 7.2 and 35.5 h that correspond to the $\alpha(t)$ curve in Fig. 1, from Eqs. (12) and (14) we obtain $C_+ = 6.9 \times 10^{-4}$ and $\kappa = 0.205$ h$^{-1}$ for Knowles et al.'s two basic kinetic parameters characterizing this particular $\alpha(t)$ curve at the molecular level. As to the KJMA Eq. (1), it does not seem merely a coincidence that with the help of two free parameters only, this mathematically simple equation is able to describe rather well the $\alpha(t)$ dependence for a great variety of protein systems and under a wide range of experimental conditions. A detailed analysis of the reason why this is so could therefore provide a new insight into the kinetics of protein aggregation. Such an analysis is however not straightforward, because while the KJMA theory predicts $n \leq 4$, as noted above, most of the aggregation experiments are characterized with $n$ values between 6 and 22. This is a clear indication that the original KJMA theory needs an appropriate modification in order to take into account the peculiarities in the evolution of the protein aggregates, e.g. the aggregate fragmentation, and in this way to become applicable to the kinetics of overall protein aggregation. A discussion on such a modification can be found in the online version of this article (as supplementary material).

**ACKNOWLEDGEMENTS:** We thank Dr. Marcus Fändrich for providing the data shown in Fig. 3, as well as Dr. Katy E. Routledge, Dr. Wei-Feng Xue and Prof. Sheena Radford for providing the data shown in Figs. 1 and 2.

## REFERENCES

1. Sunde M, Blake C. The structure of amyloid fibrils by electron microscopy and X-ray diffraction. Adv Protein Chem 1997;50:123-159.

2. Chiti F, Dobson CM. Protein misfolding, functional amyloid, and human disease, Annu Rev Biochem 2006;75:333-366.

3. Selkoe DJ. Folding proteins in fatal ways. Nature 2003;426:900-904.

4. Sawaya MR, Sambashivan S, Nelson R, Ivanova MI, Sievers SA, Apostol MI, Thompson MJ, Balbirnie M, Wiltzius JJW, McFarlane HT, Madsen AO, Riekel C, Eisenberg D. Atomic structures of amyloid cross-β spines reveal varied steric zippers. Nature 2007;447:453-457.

5. Tycko R. Molecular structure of amyloid fibrils: insights from solid-state NMR. Q Rev Biophys 2006;39:1-55.

6. Hofrichter J, Ross PD, Eaton WA. Kinetics and mechanism of deoxyhemoglobin S gelation: a new approach to understanding sickle cell disease. Proc Natl Acad Sci USA 1974;71:4864-4868.

7. Ferrone FA, Hofrichter J, Sunshine HR, Eaton WA. Kinetic studies of photolysis-induced gelation of sickle cell hemoglobin suggest a new mechanism. Biophys J 1980;32:361-377.

8. Bishop MF, Ferrone FA. Kinetics of nucleation-controlled polymerization. Biophys J 1984;46:631-644.

9. Ferrone FA, Hofrichter J, Eaton WA. Kinetics of sickle cell hemoglobin polymerization. I. Studies using temperature-jump and laser photolysis techniques. J Mol Biol 1985;183:591-610.

10. Ferrone FA, Hofrichter J, Eaton WA. Kinetics of sickle cell hemoglobin polymerization. II. A double nucleation mechanism. J Mol Biol 1985;183:611-631.

11. Jarrett JT, Lansbury PT. Seeding "one-dimensional crystallization" of amyloid: A pathogenic mechanism in Alzheimer's disease and scrapie? Cell 1993;73:1055-1058.

12. Lomakin A, Chung DS, Benedek GB, Kirschner DA, Teplow DB. On the nucleation and growth of amyloid beta-protein fibrils: detection of nuclei and quantitation of rate constants. Proc Natl Acad Sci USA 1996;93:1125-1129.

13. Lomakin A, Teplow DB, Kirschner DA, Benedek GB. Kinetic theory of fibrillogenesis of amyloid β-protein. Proc Natl Acad Sci USA 1997;94:7942-7947.

14. Serio TR, Cashikar AG, Kowal AS, Sawicki GJ, Moslehi JJ, Serpell L, Arnsdorf MF, Lindquist SL. Nucleated conformational conversion and the replication of conformational information by prion determinants. Science 2000;289:1317-1321.

15. Galkin O, Vekilov PG. Mechanisms of homogeneous nucleation of polymers of sickle cell anemia hemoglobin in deoxy state. J Mol Biol 2004;336, 43-59.

16. Galkin O, Nagel RL, Vekilov PG. The kinetics of nucleation and growth of sickle cell hemoglobin fibers. J Mol Biol 2007;365:425-439.

17. Auer S, Dobson CM, Vendruscolo M. Characterization of the nucleation barriers for protein aggregation and amyloid formation. HFSP J 2007;1:137-146.

18. Auer S, Dobson C, Vendruscolo M, Maritan A. Self-templated nucleation in peptide and protein aggregation. Phys Rev Lett 2008;101:258101.

19. Xue WF, Homans SW, Radford SE. Systematic analysis of nucleation-dependent polymerization reveals new insights into the mechanism of amyloid self-assembly. Proc Natl Acad Sci USA 2008;105:8926-8931.

20. Routledge KE, Tartaglia GG, Platt GW, Vendruscolo M, Radford SE. Competition between intramolecular and intermolecular interactions in an amyloid-forming protein. J Mol Biol 2009;389:776-786.

21. Knowles TPJ, Waudby CA, Devlin GL, Aguzzi A, Vendruscolo M, Terentjev EM, Welland ME, Dobson CM. An analytical solution to the kinetics of filament assembly. Science 2009;326:1533-1537.

22. Zhang J, Muthukumar MJ. Simulations of nucleation and elongation of amyloid fibrils. J Chem Phys 2009;130:035102.

23. Nielsen L, Khurana R, Coats A, Frokjaer S, Brange J, Vyas S, Uversky VN, Fink AL. Effect of environmental factors on the kinetics of insulin fibril formation: elucidation of the molecular mechanism. Biochemistry 2001;40:6036-6046.

24. Kim YS, Cape SP, Chi E, Raffen R, Wilkins-Stevens P, Stevens FJ, Manning MC, Randolph TW, Solomon A, Carpenter JF. Counteracting effects of renal solutes on amyloid fibril formation by immunoglobulin light chains. J Biol Chem 2001;276:1626-1633.

25. Zhu L, Zhang XJ, Wang LW, Zhou JM, Perret S. Relationship between stability of folding intermediates and amyloid formation for the yeast prion protein ure2p: a quantitative analysis of the effects of pH and buffer system. J Mol Biol 2003;328:235-254.

26. Hortschansky P, Schroeckh V, Christopeit T, Zandomeneghi G, Fändrich M. The aggregation kinetics of Alzheimer's β-amyloid peptide is controlled by stochastic nucleation. Protein Sci 2005;14:1753-1759.

27. Christopeit T, Hortschansky P, Schroeckh V, Gührs K, Zandomeneghi G, Fändrich M. Mutagenic analysis of the nucleation propensity of oxidized Alzheimer's β-amyloid peptide. Protein Sci 2005;14:2125-2131.

28. Grudzielanek S, Smirnovas V, Winter R. Solvation-assisted pressure tuning of insulin fibrillation: from novel aggregation pathways to biotechnological applications. J Mol Biol 2006;356:497-509.

29. Pedersen JS, Flink JM, Dikov D, Otzen D. Sulfates dramatically stabilize a salt dependent type of glucagon fibrils. Biophys J 2006;90:4181-4194.

30. Fändrich M. Absolute correlation between lag time and growth rate in the spontaneous formation of several amyloid-like aggregates and fibrils. J Mol Biol 2007;365:1266-1270.

31. Kashchiev D. Nucleation: Basic Theory with Applications. Oxford: Butterworth-Heinemann; 2000.

32. Carulla N, Caddy GL, Hall DR, Zurdo J, Gairi M, Feliz M, Giralt E, Robinson CV, Dobson CM. Molecular recycling within amyloid fibrils. Nature 2005;436:554-558.

33. Kunes KC, Cox DL, Singh RRP. One-dimensional model of yeast prion aggregation. Phys Rev E 2005;72:051915.

34. Kashchiev D. The kinetic approach to nucleation. Cryst Res Technol 1984;19:1413-1423.

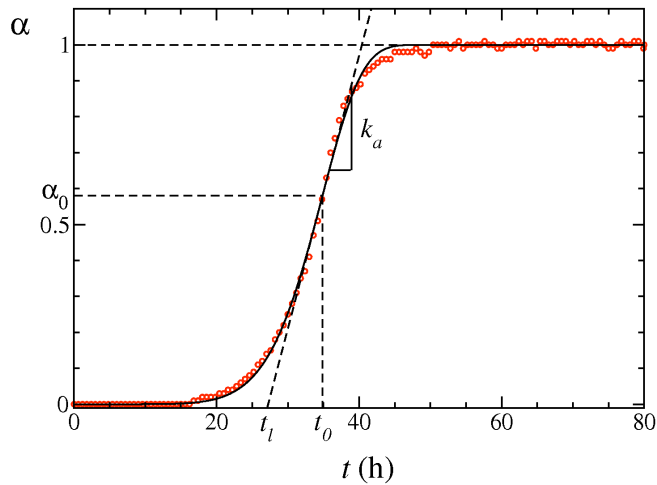

**Figure 1** Time dependence of normalized fluorescence signal: circles – experimental data of Xue et al.;[19] solid line – best fit of Eq. (1) with $\theta = 35.5$ h and $n = 7.2$. The other lines illustrate the determination of the lag time $t_l$ and the maximal aggregation rate $k_a$ with the aid of the $\alpha(t)$ inflection point coordinates $t_0$ and $\alpha_0$.

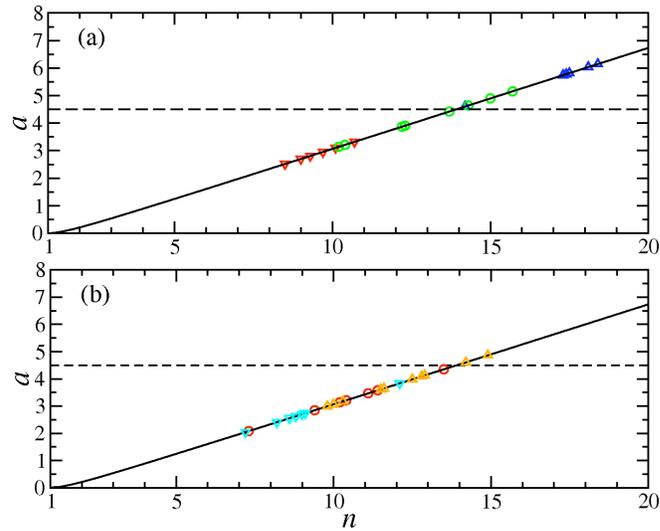

**Figure 2** Dependence of the parameter $a\,(=t_l k_a)$ on the kinetic index $n$ in overall aggregation of: (a) wild-type $\beta_2$m (up triangles), L65A $\beta_2$m (circles) and Y63A $\beta_2$m (down triangles);[20] (b) $\beta_2$m at concentrations of 9 µM (circles), 122 µM (up triangles) and 243.5 µM (down triangles).[19] The solid lines represent the $a(n)$ dependence from Eq. (9), and the dashed lines indicate Fändrich's $a = 4.5$.[30]

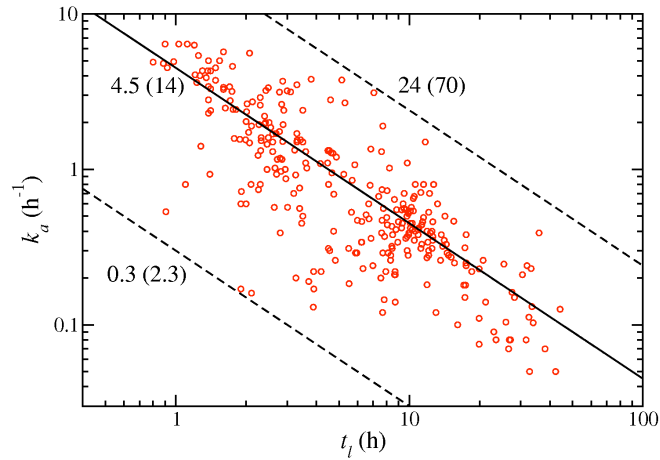

**Figure 3** Illustration of the correlation $k_a = a/t_l$ between the aggregation rate $k_a$ and the lag time $t_l$. The circles represent all data points analyzed by Fändrich,[30] the solid line corresponds to Fändrich's $a = 4.5$, and the dashed lines correspond to $a = 0.3$ and 24 (as indicated). The respective $n$ values are noted in parentheses.